\documentclass[twoside,titlepage,final,12pt]{article}
\usepackage{latexsym}
\input boxedeps
\SetRokickiEPSFSpecial
\HideDisplacementBoxes

\catcode`\@=11
\def\ps@ppt{\def\@oddhead{\qquad\textit{Top}-ology for \textit{Physics 
Today} \hfil \thepage\qquad}\def\@evenhead{\qquad\thepage \hfil \textsc{Chris Quigg} \qquad}
\def\@oddfoot{}\def\@evenfoot{}}    
\catcode`\@=12

\newcommand{\ys}{\hbox{ ys}}
\newcommand{\ps}{\hbox{ ps}}
\newcommand{\s}{\hbox{ s}}
\newcommand{\any}{\hbox{ anything}}
\newcommand{\mevcc}{\ensuremath{\hbox{ MeV}\!/\!c^2}}

\newcommand{\gevcc}{\ensuremath{\hbox{ GeV}\!/\!c^2}}
\newcommand{\gev}{\ensuremath{\hbox{ GeV}}}
\newcommand{\tev}{\ensuremath{\hbox{ TeV}}}
\newcommand{\pb}{\ensuremath{\hbox{ pb}}}

\newcommand{\HRule}{\rule{\linewidth}{1mm}}
\newcommand{\jpsi}{\ensuremath{J\!/\!\psi}}

\def\single {
                \renewcommand{\baselinestretch}{1}
                \large
                \normalsize
                }
\def\onepointfive{
                  \renewcommand{\baselinestretch}{1.5}
                  \large
                  \normalsize
                  }

\def\etal{\hbox{\it et al.}}

\def\pl#1#2#3{{\it Physics Letters }{\bf #1}, #2 (19#3)}
\def\prl#1#2#3{{\it Physical Review Letters }{\bf #1}, #2 (19#3)}
\def\rmp#1#2#3{{\it Reviews of Modern Physics }{\bf #1}, #2 (19#3)}

\def\pr#1#2#3{{\it Physical Review D\/}{\bf #1}, #2 (19#3)}
\def\arnps#1#2#3{{\it Annual Review of Nuclear and Particle Science }{\bf #1}, #2 (19#3)}
\def\np#1#2#3{{\it Nuclear Physics }{\bf #1}, #2 (19#3)}

\def\ib#1#2#3{{\it ibid. }{\bf #1}, #2 (19#3)}

\def\hepph#1{(electronic archive: hep-ph/#1)}
\def\hepex#1{(electronic archive: hep-ex/#1)}
\def\yadfiz#1#2#3#4#5#6{\frenchspacing{\it Yad. Fiz. }{\bf #1}, #2 (19#3) [English translation: 
         {\frenchspacing \it Sov. J. Nucl. Phys. }{\bf #4}, #5 (19#6)]}

\def\jetpl#1#2#3#4#5#6{\frenchspacing{\it ZhETF Pis'ma }{\bf #1}, #2 (19#3) [English translation: 
         {\frenchspacing \it JETP Lett. }{\bf #4}, #5 (19#6)]}
         
\def\phystoday#1#2#3#4{\frenchspacing{\it Phys. Today }{\bf #1}, #2 (\ifcase#3\or January\or 
         February\or March\or April\or May\or June\or July\or August\or 
         September\or October\or November\or December\fi, 19#4)}
\def\slashii#1{\setbox0=\hbox{$#1$}             
   \dimen0=\wd0                                 
   \setbox1=\hbox{\sl/} \dimen1=\wd1            
   \ifdim\dimen0>\dimen1                        
      \rlap{\hbox to \dimen0{\hfil\sl/\hfil}}   
      #1                                        
   \else                                        
      \rlap{\hbox to \dimen1{\hfil$#1$\hfil}}   
      \hbox{\sl/}                               
   \fi}                                         %
\newcommand{\met}{\ensuremath{\slashii{E}_{T}}}
\def\ltap{\mathop{\raisebox{-.4ex}{\rlap{$\sim$}} 
\raisebox{.4ex}{$<$}}}

\begin{document}
\begin{flushright}
	\textsf{FERMILAB--Pub--97/091--T \\ April 11, 1997}
\end{flushright}
\vspace*{\stretch{1}}
\HRule
\begin{center}
	{\Huge \textit{Top}-ology} \\[5mm]
	{\large Chris Quigg* } \\[5mm]
	Fermi National Accelerator Laboratory, Batavia, Illinois 60510 \\
	Department of Physics, Princeton University, Princeton, New Jersey 
	08540 \\[8mm]
	\parbox{2.9in}{Extended version of an article to appear in the May 1997 
	issue of \textit{Physics Today.}}	
\end{center}
\HRule
\vspace*{\stretch{1.5}}
		*E-mail:quigg@fnal.gov.
\newpage
\setlength{\parindent}{2ex}
\setlength{\parskip}{12pt}
\noindent
Top is a most remarkable particle, even for a quark.  A
single top quark weighs $175\gevcc$, about as much as an atom of gold.  But unlike 
the gold atom, which can be disassembled into 79 protons, 79 electrons, and 118 
neutrons, top seems indivisible, for we discern no structure
at a resolution approaching $10^{-18}\hbox{ m}$.
Top's expected lifetime of about 0.4 yoctosecond ($0.4 \times 10^{-24}\s$) 
makes it by far the most ephemeral of the quarks.  The compensation 
for this exceedingly brief life is a measure of freedom: top decays 
before it experiences the confining influence of the strong 
interaction.  In spite of its 
fleeting existence, the top quark helps shape the character 
of the everyday world.

\section*{Top Search and Discovery}
Ever since the existence of the $b$-quark was inferred from the 
discovery of the $\Upsilon$ (Upsilon) family of resonances at 
Fermilab in 1977, particle physicists have been on the lookout for its 
partner, called top.  The long search, which  
occupied experimenters at laboratories around the world, came to a 
successful conclusion in 1995 with the announcement 
that the top quark had been observed in the CDF and D\O\ experiments 
at Fermilab \cite{WWrev}.

Top is the last of the fundamental constituents of 
subnuclear matter that gauge theories of the strong, 
weak, and electromagnetic interactions and a wealth of experimental 
information have led particle physicists to expect.  Top's existence 
was required lest quantum corrections clash with the symmetries of the 
electroweak theory, leaving it internally inconsistent.  It
was signalled too by the pattern of disintegrations of the $b$-quark and 
by the characteristics of the $b$-quark's neutral--weak-current 
interactions measured in $e^{+}e^{-}$ annihilations into  $b\bar{b}$ 
pairs.  

\begin{figure}[tbh]
	\centerline{\BoxedEPSF{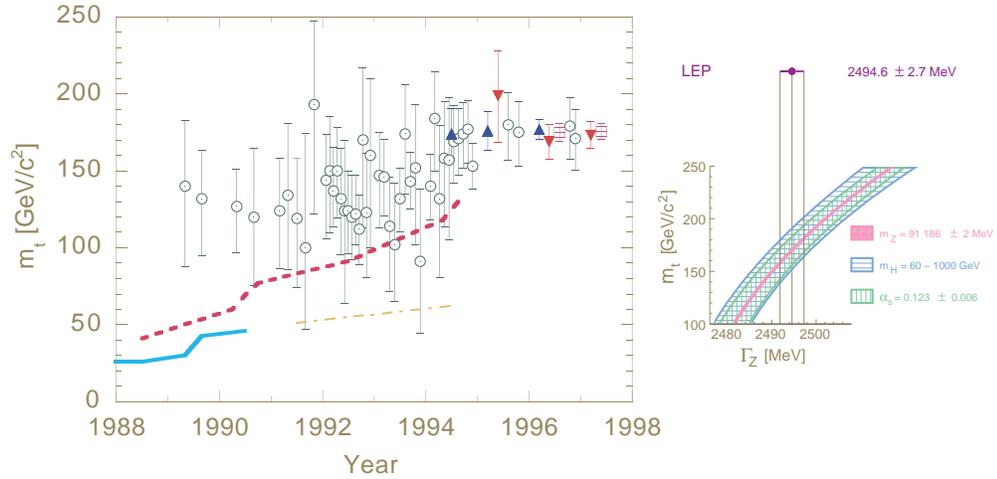  scaled 500}}
	\caption{Indirect determinations of the top-quark mass from fits to 
	electroweak observables (open circles) and 95\% confidence-level
	lower bounds on the top-quark 
	mass inferred from direct searches in $e^{+}e^{-}$ annihilations 
	(solid line) and in $\bar{p}p$ collisions, assuming that standard 
	decay modes dominate (broken line).  An indirect lower bound, derived 
	from the $W$-boson width inferred from $\bar{p}p \rightarrow 
	(W\hbox{ or }Z)+\hbox{ anything}$, is shown as the dot-dashed line.  
	Direct measurements of $m_{t}$ by the CDF (triangles) and D\O\ 
	(inverted triangles) Collaborations are shown at the time of initial evidence, 
	discovery claim, and today.  The current world 
	average from direct observations is shown as the crossed box.  For 
	sources of data, see Ref. \cite{pdg}.  \textit{Inset:} Electroweak 
	theory predictions for the width of the $Z^{0}$ boson as a function 
	of the top-quark mass, compared with the width measured in LEP  
	 experiments (Ref. \cite{zwidth}).}
	\protect\label{EWtop}
\end{figure}

Higher-order processes involving virtual top quarks are an important element 
in quantum 
corrections to the predictions the electroweak theory makes for many 
observables.  A case in point is the total decay rate, or width, of 
the $Z^{0}$ boson, which has been measured to exquisite precision at 
the CERN and SLAC $Z$ factories.  The comparison of experiment and 
theory shown in the inset to Figure \ref{EWtop} favors a top mass in 
the neighborhood of $180\gevcc$.  The top mass favored by simultaneous 
fits to many electroweak observables is shown as a function of time in 
Figure \ref{EWtop}.\marginpar{{\small \textsf{Figure \ref{EWtop}:\\ 
$m_{t}(t)$}}}

It is worth mentioning another hint that I have to confess seems more 
suggestive to me after the fact than it did before.  In 
supersymmetric unified theories of the fundamental interactions, 
virtual top quarks can drive the spontaneous breakdown of 
electroweak symmetry---provided top is very massive \cite{AGPW}.

Through the 1980s and early 1990s, direct searches continually raised the 
lower bound on the 
top mass, but produced no convincing sign of the top quark.  The most 
stringent limits came from the proton-antiproton colliders at CERN 
and Fermilab, but these 
relied on the assumption that top decays (almost) exclusively into a 
bottom quark and a real or virtual $W$ boson.  Electron-positron 
colliders could look for $e^{+}e^{-}\rightarrow t\bar{t}$ without 
assumptions about the decay mechanism, but the lower energies of 
those machines led to rather weak bounds on $m_{t}$.

By 1994, an impressive body of circumstantial evidence pointed to
the existence of a top quark with a 
mass of $175 \pm 25\gevcc$.  
Finding top and measuring its mass directly emerged as a critical test 
of the understanding of weak and electromagnetic interactions
built up over two decades.

The decisive experiments were carried out at Fermilab's Tevatron, in 
which a beam of 900-GeV protons collides with a beam of 900-GeV 
antiprotons.
Creating top-antitop pairs in sufficient numbers to claim discovery 
demanded 
exceptional performance from the Tevatron, for only one interaction in 
ten billion results in a top-antitop pair.  Observing traces 
of the disintegration of top into a $b$-quark and a $W$-boson, the 
agent of the weak interaction, required highly capable detectors and 
extraordinary attention to experimental detail.  Both the $b$-quark 
and the $W$-boson are themselves unstable, with many multibody decay 
modes.  The $b$-quark's mean lifetime is about $1.5\ps$.  It can be 
identified by a decay vertex displaced by a fraction of a millimeter 
from the production point, or by the low-momentum electron or muon from the 
semileptonic decays $b \rightarrow ce\nu$, $b \rightarrow c\mu\nu$, 
each with branching fraction about 10\%.  The $W$ boson decays after 
only $0.3\ys$ on average into $e\bar{\nu}_{e}$, 
$\mu\bar{\nu}_{\mu}$, $\tau\bar{\nu}_{\tau}$, or a quark and 
antiquark (observed as two jets of hadrons), 
with probabilities 1/9, 1/9, 1/9, and 2/3.  The characteristic modes 
in which $t\bar{t}$ production can be sought are shown with their 
relative weights in Table \ref{canaux}.
\marginpar{{\small \textsf{Table \ref{canaux}: \\ Search modes}}}
Dilepton events ($e\mu,$ $ee$, and $\mu\mu$) are produced primarily 
when both $W$ bosons decay into $e\nu$ or $\mu\nu$.  Events in the 
lepton + jets channels $(e,\mu + \hbox{jets})$ occur when one $W$ 
boson decays into leptons and the other decays through quarks into 
hadrons.

\begin{figure}[tbh]
	\centerline{\BoxedEPSF{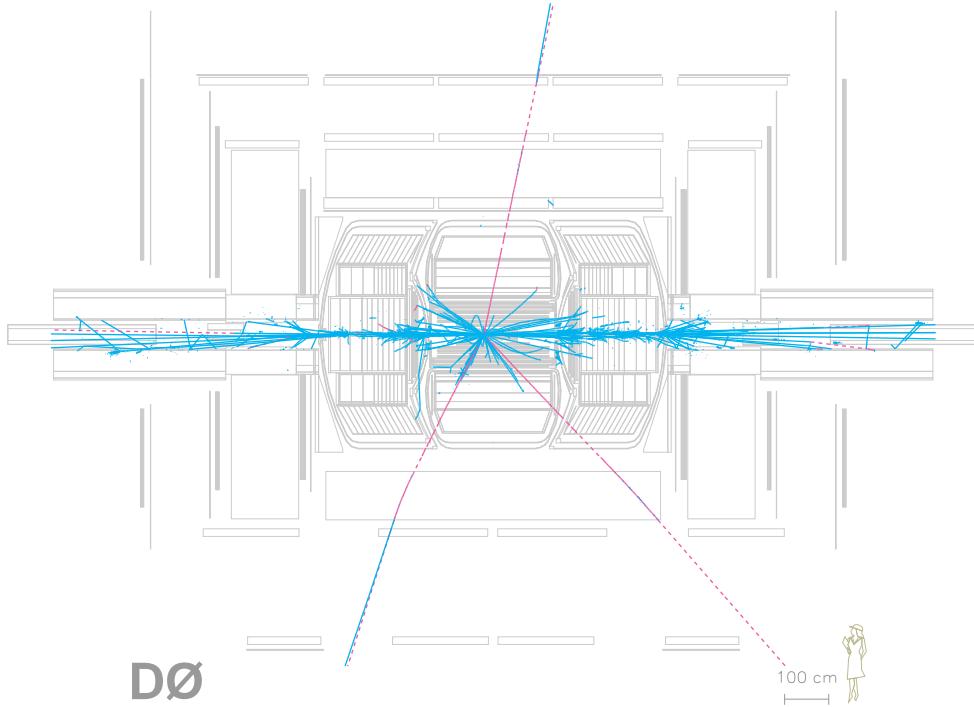  scaled 500}}
	\caption{Simulation of a top-antitop event produced in a 2.0-TeV 
	proton-antiproton collision in the upgraded D\O\ 
	detector, which will operate at the Fermilab Tevatron starting in 
	1999.  The beam particles entered horizontally and collided at the center 
	of the picture.
    The light blue lines are the trajectories of charged hadrons, electrons and
    positrons produced in the collision; the pink lines represent muons.
    In this event, both $W$-bosons produced in top decays subsequently
    produced high energy muons (the tracks at upper center and lower right).  
    A third muon (lower left) originated in the decay of a $b$-quark; its 
    lower momentum can be inferred from the noticeable curvature of its track 
    in the magnetized-iron section of the detector.  
    The event was generated using the ISAJET Monte Carlo of F.E. Paige and 
    S. Protopopescu and the detector was simulated using the GEANT
    package from the CERN program library.  I thank John Womersley for 
    supplying this figure.}
	\protect\label{worm}
\end{figure}

Another challenge to experiment is the complexity of events in 
high-energy $\bar{p}p$ collisions.  The top and antitop are typically 
accompanied by scores of other particles.  Figure \ref{worm} shows a 
\marginpar{{\small \textsf{Figure \ref{worm}: \\ D\O\ Simulation}}} 
simulated $t\bar{t}$ event in the D\O\ detector.  The only 
characteristic features evident to the eye are the penetrating muons
 near the top center and bottom right, which suggest two $W 
\rightarrow \mu\nu$ decays, and the low-momentum muon at lower 
left.  Separating the top-quark sheep from the 
goats is not for the faint of heart!

Each detector is an intricate apparatus operated by an 
international collaboration of about 450 physicists.  The 
tracking
devices, calorimeters, and surrounding iron for muon identification 
occupy a volume about three stories high and weigh about 5000 tons.  
The Collider Detector at Fermilab (CDF), a magnetic detector with 
solenoidal geometry, profited from its high-resolution silicon vertex 
detector (SVX) to tag $b$-quarks with good efficiency.  The D\O\ Detector 
(D-Zero) has no central magnetic field, emphasizing instead calorimetric 
measurement of the energies of produced particles.

\begin{figure}[tbh]
	\centerline{\BoxedEPSF{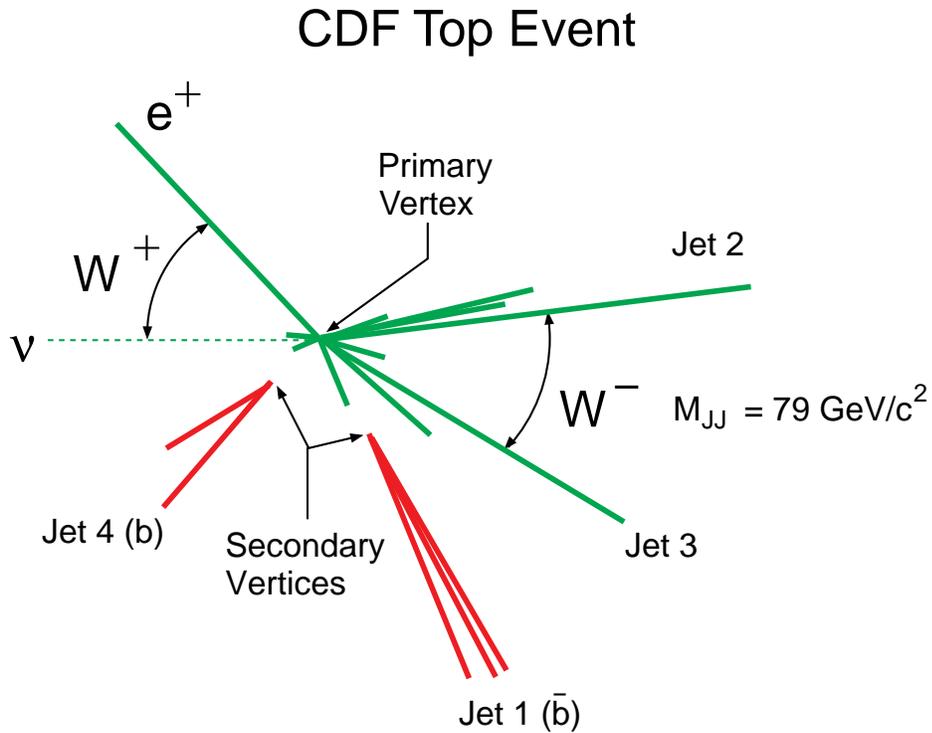  scaled 700}}
	\caption{Candidate event for top-antitop production, as seen by CDF's 
	silicon vertex detector at the Tevatron.  Both top quarks decay at 
	the $p\bar{p}$ collision vertex into a $W$-boson plus a bottom 
	quark.  The $W^{+}$ decays to $e^{+}$ plus an invisible neutrino, and 
	the $W^{-}$ decays into a quark and antiquark that show up as jets of 
	hadrons.  Each bottom quark becomes a $B$ meson that travels a few 
	millimeters from the production vertex before its decay creates a 
	hadron jet.  Many extraneous tracks are not shown.}
	\protect\label{CDFSVX}
\end{figure}

The first evidence for top was presented in April 1994 by the CDF 
Collaboration, led by Bill Carithers of Lawrence Berkeley Laboratory 
and Mel Shochet of the University of Chicago \cite{cdf1}.
\marginpar{{\small \textsf{Figure \ref{CDFSVX}: \\ CDF SVX event}}}
In a sample of 19.3 events per picobarn of cross section 
($19.3\pb^{-1}$), CDF found 12 events consistent with 
either two $W$ bosons, or a $W$ boson and at least one $b$-quark.  
One of the $e\mu$ candidates, shown in Figure \ref{CDFSVX}, shows the 
power of the SVX to resolve a $b$-decay vertex just 0.3~mm 
from the interaction point.  Although the sample lacked the 
statistical weight needed to claim discovery, the event 
characteristics were consistent with the $t\bar{t}$ interpretation, 
with a top mass of $174 \pm 10 ^{+13}_{-12}\gevcc$.  A few months 
later, the D\O\ Collaboration reported an excess of candidates (9 
events with an expected background of $3.8 \pm 0.9$) in a 
13.5-pb$^{-1}$ sample \cite{dzero1}.

The discovery was not far behind.  By February 1995, both groups had 
quadrupled their data sets.  The CDF Collaboration, now led by 
Carithers and Giorgio Bellettini of the University of Pisa, found 6 
dilepton candidates with an anticipated background of $1.3 \pm 0.3$ 
events, plus 37 $b$-tagged events containing a $W$-boson and at least 
three jets \cite{CDF}.  The D\O\ 
Collaboration, with Paul Grannis of Stony Brook and Hugh Montgomery of 
Fermilab as spokespersons, reported 17 top candidates with an expected 
background of $3.8 \pm 0.6$ \cite{Dzero}.  Taken 
together, the populations and characteristics of different event 
classes provided irresistible evidence for a top quark with a mass in 
the anticipated region: $176\pm 8 \pm 10\gevcc$ for 
CDF, and $199^{+19}_{-21}\pm 22\gevcc$ for D\O.  The 
top-antitop production rate is in line with theoretical predictions.
\marginpar{{\small \textsf{Box: The Third \\ \hphantom{Box:} Generation}}}

Today, with the event samples approximately doubled again, the top 
mass is measured as $176.8 \pm 6.5\gevcc$ by CDF and $173.3 \pm 
8.4\gevcc$ by D\O\, for a world average of $175.5 \pm 5.1\gevcc$. 

Now that we have the top quark, what do we do with it?

\section*{The Top Quark and the $W$ Boson}
The influence of virtual top quarks was the basis for the 
expectations for the top-quark mass from precision measurements of 
electroweak observables.  As $m_{t}$ becomes known more precisely from 
direct measurements, it will be possible to compare predictions that 
depend sensitively on $m_{t}$ with new observations.  Among the most 
incisive will be the comparison of the $W$-boson mass with theoretical 
calculations.  

\begin{figure}[tbh]
	\centerline{\BoxedEPSF{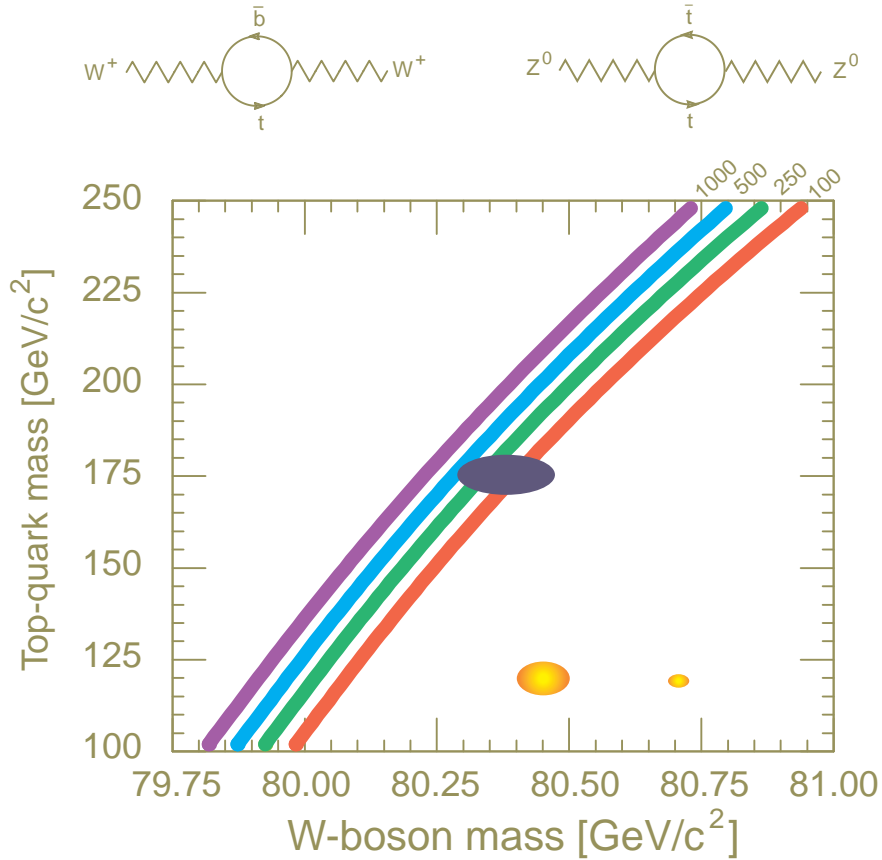  scaled 750}}
	\caption{Correlation between the top-quark mass and the $W$-boson 
	mass in the standard electroweak theory.  From left to 
	right, the bands correspond to Higgs-boson masses of $1000,\:500,\: 
	250, \hbox{ and }100\gevcc$.  The thickness of the bands expresses 
	the effect of plausible variations in the value of 
	$\alpha(M_{Z})$.  The dark region is the 
	one-standard-deviation error ellipse from the current world 
	averages, $m_{t}=175.5 \pm 5.1\gevcc$ and $M_{W} = 80.38 \pm 
	0.09\gevcc$.  
	Also shown are the one-standard-deviation error ellipses for 
	precisions expected in the future: ($\delta M_{W}=50\mevcc, \;\delta 
	m_{t}=5\gevcc$) and ($\delta M_{W}=20\mevcc,\; \delta 
	m_{t}=2\gevcc$).  Examples of the heavy-quark loops that give rise to 
	$\Delta\rho$ are shown at the top of the figure.}
	\protect\label{MWmt}
\end{figure}

The $W$-boson mass is given as
\begin{equation}
	M_{W}^{2} = M_{Z}^{2} (1 - \sin^{2}\theta_{W})(1 + \Delta\rho),
	\label{rho}
\end{equation}
where $M_{Z}$ is the mass of the $Z^{0}$ boson, 
$\sin^{2}\theta_{W}\approx 0.232$ is the weak mixing parameter, and 
$\Delta\rho$ represents quantum corrections.  The most important of 
these are shown in Figure \ref{MWmt}.  The inequality of the $t$- 
and $b$-quark masses violates weak-isospin symmetry and results in
\begin{equation}
	\Delta \rho = 3G_{F}m_{t}^{2}/8\pi^{2}\sqrt{2} + \ldots,
	\label{deltarho}
\end{equation}
where the unwritten terms include a logarithmic dependence upon the 
mass of the Higgs boson, the hitherto undetected agent of electroweak 
symmetry breaking.

Predictions for $M_{W}$ as a function of the top-quark mass are shown 
in Figure \ref{MWmt}
\marginpar{{\small \textsf{Figure \ref{MWmt}: \\ $M_{W}(m_{t})$}}}
for several values of the Higgs-boson mass \cite{tatsu}.  Current 
measurements are consistent with the electroweak theory, but do not 
yet provide any precise hints about the mass of the Higgs boson.  The 
uncertainty on the world-average $M_{W}$ has now reached about $100\mevcc$.  
An uncertainty of $\delta M_{W}=50\mevcc$ seems a realistic 
possibility both at the Tevatron and at CERN's LEP200, where 
observations of the reaction $e^{+}e^{-}\rightarrow W^{+}W^{-}$ near 
threshold began in 1996.  Improving $\delta m_{t}$ below $5\gevcc$ will 
then make for a demanding test of the electroweak theory that should 
yield interesting clues about the Higgs-boson mass.  Over the next decade, 
it seems 
possible to reduce $\delta m_{t}$ to $2\gevcc$ at Fermilab and $\delta 
M_{W}$ to about $20\mevcc$ at the Tevatron and LEP200.  That will set 
the stage for a crucial test of the electroweak theory when (and if) the Higgs 
boson is discovered.

\section*{Is It Standard Top?}
The top-quark discovery channels listed in Table \ref{canaux} all 
arise from the production of top-antitop pairs, and all contain a 
$b\bar{b}$ pair.  We expect that top decays other than the observed 
$t \rightarrow bW^{+}$ mode are strongly suppressed.  Unless the $t 
\rightarrow bW^{+}$ rate is unexpectedly small, which could occur 
if top had a large coupling to a more massive, fourth-generation 
$b^{\prime}$, the decays $t \rightarrow (s\hbox{ or }d)W^{+}$ should 
be extremely rare.  It is important to test this expectation by 
looking for the rare decays directly, or by comparing the number of 
observed (0, 1, and 2) $b$-tags in a $t\bar{t}$ sample with 
expectations derived from the measured efficiency for $b$-tagging.  The CDF 
Collaboration has used the tagging method to show that $t \rightarrow 
bW$ accounts for $99\pm 29\%$ of all $t \rightarrow 
W+\hbox{ anything}$ decays \cite{joei}.

Top pairs are produced in $\bar{p}p$ collisions through the strong interaction.  
A single top can be produced together with an antibottom in processes 
that involve the weak interaction.  The elementary process $u\bar{d} 
\rightarrow \hbox{virtual }W^{+} \rightarrow t\bar{b}$ may in time 
give us an excellent measurement of the strength of the $Wt\bar{b}$ 
coupling.

In some supersymmetric models, top can be produced in the decays of 
heavy superpartners and can itself decay into lighter superpartners.  
This possibility encourages the careful comparison of the top-bearing 
events with conventional expectations, and emphasizes the importance 
of precision determinations of the top production cross section.

The rapid decay of the top quark means that there is no time for the 
\marginpar{{\small \textsf{Box: The Brief, \\ \hphantom{Box:} Happy 
Life\ldots}}}
formation of top mesons or top baryons.  Accordingly, the spin 
orientation of the top quark at the moment of its production is 
reflected, without dilution, in the decay angular distribution of its 
decay products.  The correlation between the spin of the top and 
antitop produces distinctive patterns in the structure of events that 
will enable us to probe the character of the $t \rightarrow bW^{+}$ 
transition.

If top's weak interactions are as expected, top decay is an excellent source 
of longitudinally polarized $W$ bosons.  A fraction 
$(1+2M_{W}^{2}/m_{t}^{2})^{-1} \approx 70\%$ of the $W$ bosons in top decay will 
be longitudinally polarized.  That polarization is reflected in the 
decay angular distribution of the electrons and muons from $W$ decay.  
The longitudinal $W$s are interesting in their own right: as creatures 
of electroweak symmetry breaking, they may be particularly sensitive to 
new physics.    

Because top is so massive, many decay channels may be open to it, 
in addition to the signature $t \rightarrow bW^{+}$ mode.  The decay into 
a $b$-quark and a charged spin-zero particle $P^{+}$
may occur in multi-Higgs generalizations of the electroweak 
theory, in supersymmetric models, and in technicolor models.  The 
decay rate for $t \rightarrow bP^{+}$ is similar to 
the $t \rightarrow bW^{+}$ rate, because both decays are semiweak.  If 
the $t\bar{t}$ production rate were measured to be smaller than 
predicted by QCD, that would hint at nonstandard decays---and new 
physics.  The lifetime of $P^{+}$, typically about $10^{-21}\s=1
\hbox{ zeptosecond}$, is far too short for it to be observed as a 
short track.  $P^{+}$ might be recognized from its decays into 
heavy quarks or into $\tau \nu_{\tau}$.  The general lesson is that top 
decays have the capacity 
to surprise. 

\section*{Top and Electroweak Symmetry Breaking}
What sets the masses of the fundamental fermions and bosons?
In the standard electroweak theory, the Higgs boson gives masses to the gauge 
bosons $W^{\pm}$ and $Z^{0}$, and to the quarks and leptons.  The 
mechanisms are linked---both arise through the breaking of 
electroweak symmetry---but they are logically distinct.  While the 
$W^{\pm}$ and $Z^{0}$ masses are predicted in terms of the coupling 
constants and the weak mixing parameter, every fermion mass is set by 
a separate Yukawa coupling.  The mass of fermion $f$ is
\begin{equation}
	m_{f} = \zeta_{f}\frac{v}{\sqrt{2}},
	\label{mf}
\end{equation}
where $v/\sqrt{2} = (2G_{F}\sqrt{2})^{-1/2} \approx 176\gev$
is the vacuum expectation value of the Higgs field 
\cite{GT}.  The Yukawa couplings range from $\zeta_{e} \approx 3 \times 
10^{-6}$ for the electron to $\zeta_{t}\approx 1$ for top.  Within the 
electroweak theory, we do not know the origin of these numbers and we 
haven't a clue how to calculate them.

Top's great mass suggests that top stands apart from the other quarks and leptons.  
Does $\zeta_{t}\approx 1$ mean that top is special, or that it is the only 
fermion with a normal mass?  We don't yet know the answer.  We expect 
that experiments at CERN's Large Hadron Collider, which will explore 
14-TeV proton-proton collisions beginning around the year 2006, will reveal the 
mechanism of electroweak symmetry breaking and complete our 
understanding of the gauge-boson masses.  But what of the fermion 
masses?  My instinct is that top's large mass means that both 
questions will be answered by experiments that probe the natural 
scale of electroweak symmetry breaking.

This is speculation, but it is certain that the discovery of top opens 
a new window on electroweak symmetry breaking.  The Higgs mechanism of 
the standard electroweak theory is the relativistic generalization of 
the Ginzburg--Landau phenomenology of the superconducting phase 
transition.  Some attempts to improve the electroweak theory and make 
it more predictive seek to emulate the Bardeen--Cooper--Schrieffer 
theory of superconductivity.  Resonances that decay into $t\bar{t}$ 
are natural consequences of these dynamical schemes.  The possibility 
of new sources of $t\bar{t}$ pairs makes it urgent to test how 
closely top production conforms to standard (QCD) expectations.

Two classes of models have received considerable attention in the 
context of the heavy top quark.  In the first, called technicolor, a new
interaction analogous to the QCD of the familiar strong interactions
becomes strong at low energies and forms a 
technifermion condensate that breaks chiral symmetry and gives masses 
to the gauge bosons.  A generalization, extended technicolor, allows the 
fermions to acquire mass through new interactions with 
the technifermion condensate.  In the second class of models, 
called topcolor, a 
new interaction drives the formation of a top condensate akin to 
Cooper pairs. The top condensate hides 
the electroweak symmetry and gives masses to the ordinary fermions.  
Top-condensate models and technicolor both imply the 
existence of color-octet resonances that decay into $t\bar{t}$, for 
which the natural mass scale is a few hundred$\gevcc$.  We are led to 
ask: Is there a resonance in $t\bar{t}$ production?  How is it made?  
How else does it decay?

In the technicolor picture, which has been elaborated recently by 
Estia Eichten and Ken Lane \cite{ELTC}, a color-octet analogue of 
the $\eta^{\prime}$ meson, called $\eta_{T}$, is produced in 
gluon-gluon interactions.  The sequence $gg \rightarrow \eta_{T} 
\rightarrow (gg, t\bar{t})$ leads to distortions of the $t\bar{t}$ 
invariant-mass distribution, and of the two-jet invariant-mass 
distribution, but has a negligible effect on the $b\bar{b}$ 
invariant-mass distribution.  

In the topcolor picture explored by Chris Hill and Stephen Parke 
\cite{TCHP}, a massive vector ``coloron'' can be produced in 
quark-antiquark interactions.  The coloron decays at comparable rates 
into $t\bar{t}$ and $b\bar{b}$ and can appear as a broad resonance peak in 
both channels.  There is no particular reason to expect a distortion 
of the invariant-mass spectrum of two jets that do not contain heavy 
quarks.

If an enhancement were seen in the $t\bar{t}$ channel, we would want to 
study the $t\bar{t}$ mass spectrum at 
different energies.  At the Tevatron, about 90\% of top-pair production 
occurs in quark-antiquark collisions.  At the much higher energy of 
the LHC, gluon-gluon collisions occur for about 90\% of the top 
pairs.  The LHC's large rate of $gg$ collisions would 
dramatically increase the contribution of $\eta_{T}$ relative to the 
coloron.
\section*{Top Matters!}
It is popular to say that 
top quarks were produced in great numbers in the 
fiery cauldron of the Big Bang some fifteen billion years ago, disintegrated 
in the merest fraction of a 
second, and vanished from the scene until my colleagues learned to create 
them in the Tevatron.  That 
would be reason enough to care about top: to learn how it helped sow the 
seeds for the primordial universe that evolved into our world of diversity 
and change.  But it is not the whole story; it invests the top quark with a 
remoteness that veils its importance for the everyday world.

The real wonder is that here and now, every minute of every day, the top 
quark affects the world around us.  Through the uncertainty principle of 
quantum mechanics, top quarks and antiquarks wink in and out of an 
ephemeral presence in our world.  Though they appear virtually, fleetingly, 
on borrowed time, top quarks have real effects.

Quantum effects make the coupling strengths of the fundamental 
interactions---appropriately normalized analogues of the 
fine-structure constant $\alpha$---vary with the energy scale on 
which the coupling is measured.  The fine-structure constant itself 
has the familiar value $1/137$ in the low-energy (or long-wavelength) 
limit, but grows to about $1/129$ at the mass of the $Z^{0}$ boson, 
about $91\gevcc$.  Vacuum-polarization effects make the effective 
electric charge increase at short distances or high energies.

\begin{figure}[tbh]
	\centerline{\BoxedEPSF{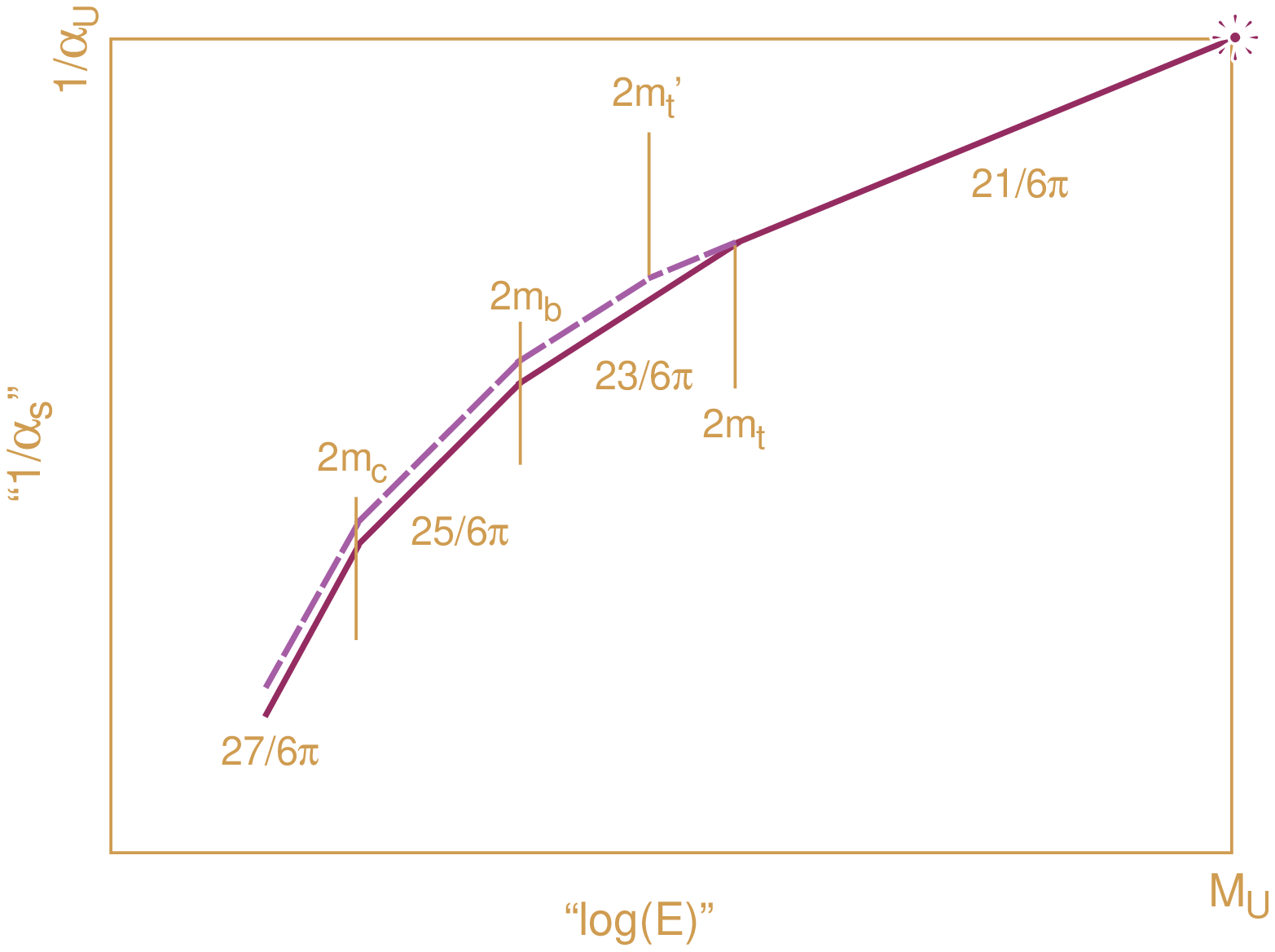  scaled 750}}
	\caption{Two evolutions of the strong coupling constant 
	$\alpha_{s}$.  A smaller value of the top-quark mass leads to a 
	smaller value of $\alpha_{s}$.}
	\protect\label{fig4}
\end{figure}

In unified theories of the strong, weak, and electromagnetic 
in\-ter\-ac\-tions, all the  coupling ``constants'' take on a 
common value, $\alpha_U$, at some high energy, $M_U$.  If we adopt the 
point of view that
$\alpha_{U}$ is fixed at the unification 
scale, then the mass of the top quark is encoded in the value of 
the strong coupling $\alpha_s$ that we 
experience at low energies \cite{su5}.  Assuming three generations of quarks and 
leptons, we  evolve $\alpha_s$ downwards in energy from the 
unification scale \cite{GQW}.
The leading-logarithmic behavior is given by
\begin{equation}
1/\alpha_s(Q) = 1/\alpha_U + \frac{21}{6\pi}\ln(Q/M_U)\;\; ,
\end{equation} for $M_U > Q > 2 m_t$.  The positive coefficient 
$+21/6\pi$ means that the strong coupling constant 
$\alpha_{s}$ is smaller at high energies than at low energies.  This 
behavior---opposite to the familiar behavior of the electric 
charge---is the celebrated property of asymptotic freedom.
In the interval between $2m_t$ and $2m_b$, the 
slope $(33-2n_{\!f})/6\pi$ (where $n_{\!f}$ is the number of active quark 
flavors) steepens to $23/6\pi$, and then increases by 
another $2/6\pi$ at every quark threshold.  At the boundary $Q=Q_n$ 
between effective field theories with $n-1$ and $n$ active flavors, the 
coupling constants $\alpha_s^{(n-1)}(Q_n)$ and $\alpha_s^{(n)}(Q_n)$ must 
match.  This behavior is 
shown by the solid line in Figure \ref{fig4}.
\marginpar{{\small \textsf{Figure \ref{fig4}: \\ $1/\alpha_{s}$ evolution}}}

The dotted line in Figure \ref{fig4} shows how the evolution of 
$1/\alpha_s$ changes if the top-quark mass is reduced.  A smaller top 
mass means a larger low-energy value of $1/\alpha_s$, so a smaller 
value of $\alpha_s$.  

Neglecting the tiny ``current-quark'' masses of the up and down 
quarks, the scale parameter $\Lambda_{\hbox{\footnotesize QCD}}$ is the 
only mass parameter in QCD.  It determines the scale of the 
confinement energy that is the dominant contribution to the proton mass. 
To a good first approximation, 
\begin{equation}
	M_{\hbox{{\footnotesize proton}}} \approx C \Lambda_{\hbox{{\footnotesize QCD}}},
	\label{lattice}
\end{equation}
where the constant of proportionality $C$ is calculable using 
techniques of lattice field theory.
 
To discover the dependence of $\Lambda_{\hbox{{\footnotesize QCD}}}$ upon the top-quark 
mass, we calculate $\alpha_s(2m_t)$ 
evolving up from low energies and down from the unification scale, and match:
\begin{equation}
1/\alpha_U  +  {\displaystyle \frac{21}{6\pi}}\ln(2m_t/M_U) =  
 1/\alpha_s(2m_c) - {\displaystyle \frac{25}{6\pi}}\ln(m_c/m_b) 
 -{\displaystyle \frac{23}{6\pi}}\ln(m_b/m_t)   .
 \end{equation} Identifying
 \begin{equation}
 1/\alpha_s(2m_c) \equiv {\displaystyle\frac{ 
 27}{6\pi}}\ln(2m_c/\Lambda_{\hbox{{\footnotesize QCD}}})\;  ,
\end{equation}  we find that 
\begin{equation}
	\Lambda_{\hbox{{\footnotesize QCD}}}=e^{\displaystyle -6\pi/27\alpha_U} 
	\left(\frac{M_U}{1 \gev}\right)^{\!21/27} 
	\left(\frac{2m_t\cdot 2m_b\cdot 2m_c}{1\gev^{3}}\right)^{\!2/27}\gev \;\; .
	\label{blank}
\end{equation}  

We conclude that, in a simple unified theory,
\begin{equation}
	\frac{M_{\hbox{{\footnotesize proton}}}}{1\gev} \propto 
	\left(\frac{m_t}{1\gev}\right)^{2/27} \;\; .
	\label{amazing}
\end{equation}
This is a wonderful result.  Now, we can't use it to
compute the mass of the top quark, 
because we don't know the values of $M_{U}$ and $\alpha_{U}$, and 
haven't yet calculated precisely the constant of proportionality 
between the proton mass and the QCD scale parameter.  Never mind!  The 
important lesson---no surprise to any twentieth-century physicist---is
that the microworld does determine the behavior 
of the quotidian.  We will fully understand 
the origin of one of the most important parameters in the everyday 
world---the mass of the proton---only by knowing the properties of the 
top quark  \cite{18params}.  
\section*{Top Priorities}
Like the end of many a scientific quest, the discovery of top marks a 
new opening \cite{snowmass}.  The first priority, already well advanced,
is to continue refining  the measurements of 
the top mass.  It is now possible to begin asking how precisely 
top fits the profile of anticipated properties in its production and 
decay.  Because of top's great mass, its decay products may include 
unpredicted---or at least undiscovered---new particles.  A very 
interesting development would be the observation of resonances in 
top-antitop production that would give new clues about the breaking of 
electroweak symmetry.  On the theoretical front, the large mass of 
top encourages us to think that the two problems of mass may be 
linked at the electroweak scale.

For the moment, the direct study of the top quark belongs to the 
Tevatron.  Early in the next century, samples twenty times greater than 
the current samples should be in hand, thanks to the increased event rate made 
possible by Fermilab's Main Injector and upgrades to CDF and D\O.  
Boosting the Tevatron's energy to $1\tev$ per beam will increase the 
top yield by nearly 40\%.
Further enhancements to Fermilab's accelerator complex are under study.  
A decade from now, the Large Hadron Collider at CERN will 
produce tops at more than ten thousand times the rate of the 
discovery experiments.  Electron-positron linear colliders or muon 
colliders may add new 
opportunities for the study of top-quark properties and dynamics.  
In the meantime, the network of understanding 
known as the standard model of particle physics links the properties 
of top to many phenomena to be explored in other experiments.

According to the cockroach theory of stock market analysis (``You 
never see just one''), there is never a single piece of good news or 
bad news.  In physics, one discovery often leads to others.  Top 
opens a new world---the domain of a very heavy fermion---in which the 
strange and wonderful may greet us.
\newpage
\section*{Acknowledgments}
I am grateful to Bob Cahn, Sekhar Chivukula, Herb Greenlee, Debbie 
Harris, Jim Hartle, Joe Incandela, Ken Lane, Kate Metropolis, David 
Pines, Liz Simmons, Stew Smith, Erich Vogt, Scott Willenbrock, Steve 
Wimpenny, Brian Winer, Bruce Winstein, and John Womersley for 
insightful comments on the manuscript.

Fermilab is operated by Universities Research Association, Inc., under
contract DE-AC02-76CHO3000 with the United States Department of Energy.  
I thank the Aspen Center for Physics for warm hospitality.  

\newpage
\nonfrenchspacing

\newpage
\noindent\HRule
\vspace{-24pt}
\section*{\textsf{Box: The Third Generation}}
The possibility that CP violation arises from complex elements of the 
quark mass matrix, for theories with at least three generations, was raised by
M. Kobayashi and T. Maskawa, \textit{Prog. Theoret. Phys. (Kyoto)} 
\textbf{49,} 652 (1973).  In the following year, the discovery of 
the \jpsi\ family of resonances by Samuel C.\ C.\ Ting's team at 
Brookhaven National Laboratory and by Burton Richter and collaborators 
at the Stanford Linear Accelerator Center completed the second 
generation of quarks and leptons.  The \jpsi\ states proved to be 
resonances of a charmed quark and charmed antiquark when mesons containing 
a single charmed quark were observed by the SLAC--Berkeley team [G. 
Goldhaber, \etal, \prl{37}{255}{76}; I. Peruzzi, \etal, 
\ib{37}{569}{76}].  The new charmed quark joined the three classical
quarks in two pairs (up, down; charm, strange) that matched the pattern of leptons 
(electron neutrino, electron; muon neutrino, muon) known since the 
early 1960s.

In 1975, Martin Perl and collaborators [\prl{35}{1489}{75}] discovered 
the $\tau$ lepton in electron-positron annihilations in the SPEAR storage 
ring at the Stanford Linear Accelerator Center. In a sample of about 
36,000 events, they found 64 that consisted of a muon and electron of 
opposite charges, plus at least two undetected particles.
The existence of the tau neutrino is inferred from the 
undetected (or ``missing'') energy of tau decay, much as the continuous 
electron energy spectrum in beta decay led Pauli to postulate the 
electron (anti)neutrino.  The tau neutrino has not yet been detected 
directly.  A tau neutrino that interacts in matter and materializes 
into a tau lepton is the hoped-for signature in a new generation of 
neutrino-oscillation searches.

	The discovery in 1977 new family of heavy mesons was the first 
	indication for a fifth quark, the $b$ (bottom, or 
	beauty), with a mass $m_{b}\approx 5\gevcc$ and charge $-1/3$.  The 
	$\Upsilon(9.46\gevcc)$ and two of its excitations were first observed by 
	Leon Lederman and his collaborators at Fermilab in the 
	reaction $p+(\hbox{Cu,Pt})\rightarrow \mu^{+}\mu^{-}+\any$ [S. W. 
	Herb, \etal, \prl{39}{252}{77}].  The $\Upsilon$ family was quickly 
	identified as a set of levels of a $b$-quark bound to a $b$-antiquark.  
	Comparison of the $b\bar{b}$ spectrum with the charmonium (\jpsi) 
	spectrum showed that the interquark force was independent 
	of the flavor of the quarks, as expected from quantum chromodynamics.  
	Hadrons containing a single $b$-quark were identified in due course in 
	the CLEO Detector at the Cornell Electron Storage Ring [S. Behrends, \etal\ 
	(CLEO Collaboration), \prl{50}{881}{83}].  The electroweak theory 
	predicts large CP-violating effects in certain $B$-meson decays.  The 
	search for these effects is a primary motivation for $B$ Factories 
	and other high-statistics $B$ experiments.
	
	Studies of $Z^{0}$ production and decay in electron-positron 
	annihilations demonstrate that there are three species of light 
	neutrinos.  The invisible decay rate of the $Z^{0}$ is determined by 
	subtracting the measured rates for decays into quarks and charged 
	leptons from the total $Z^{0}$ decay rate.  The invisible rate is 
	assumed to arise from decays into $N_{\nu}$ species of 
	neutrino-antineutrino pairs, each contributing the rate given by the 
	standard model.  Since there are only three light neutrinos, we 
	conclude that there are three ordinary generations of quarks and 
	leptons.

The top quark was found in collisions of 900-GeV protons on 900-GeV 
antiprotons at Fermilab in 1995 by the CDF and D\O\ Collaborations.

\begin{table}[h!]
	\begin{center}
		{\textsc{Quarks and leptons of the third generation.}\\[6pt]}
\onepointfive
		\begin{tabular}{|cccc|}
			\hline
			Quark & Charge & Mass & Mean Life \\ 
			\hline
			$t$ & $+2/3$ & $\sim 175\gevcc$ & $\sim 0.4\ys$ (?) \\
			\hline
			$b$ & $-1/3$ & $\sim 4.7\gevcc$ & $\sim 1.5\ps$   \\
			\hline
			\hline
			Lepton & Charge & Mass & Mean Life  \\
			\hline
			 $\nu_{\tau}$ & $0$ & $< 24\mevcc$ & $\cdots$ \\
			\hline
			$\tau$ & $-1$ & $1777.0\mevcc$ & $\sim 0.3\ps$ \\
			\hline
		\end{tabular}
		\protect\label{}
	\end{center}
\end{table}
\vspace{-24pt}
\noindent\HRule
\single
\newpage
\noindent\HRule
\vspace{-24pt}
\section*{\textsf{Box: The Brief, Happy Life of the Top Quark}}
The dominant decay of a heavy top quark is into a bottom quark and a 
$W$-boson.  This process is called \emph{semiweak,} because the rate is 
proportional to only one power of the Fermi constant $G_{F}$, whereas 
familiar weak processes like $\beta$-decay occur with rates 
proportional to $G_{F}^{2}$.  The top-quark decay rate is approximately \cite{Gamt}
\begin{eqnarray}
	\Gamma(t \rightarrow bW^+) & = & 
	\frac{G_FM_W^2}{8\pi\sqrt{2}}\frac{|V_{tb}|^{2}}{m_t^3}
	\left[\frac{(m_t^2-m_b^2)^2}{M_W^2}+m_t^2+m_b^2-2M_W^2\right] \nonumber\\
	 & & \times	\sqrt{[m_t^2-(M_W+m_b)^2][m_t^2-(M_W-m_b)^2]}.\nonumber
	\label{tdk}
\end{eqnarray}
Here $m_{t}$, $m_{b}$, and $M_{W}$ are the masses of top, bottom, and 
the $W$-boson, and $V_{tb}$ measures the strength of the $t 
\rightarrow bW^{+}$ coupling.  To the extent that the $b$-quark mass 
is negligible, 
the decay rate can be recast in the form
\begin{displaymath}
	\Gamma(t \rightarrow bW^+) = \frac{G_F m_t^3}{8\pi\sqrt{2}}
	|V_{tb}|^{2} \left(1 - \frac{M_{W}^{2}}{m_{t}^{2}}\right)^{\!\!2}
	\left(1 + \frac{2M_{W}^{2}}{m_{t}^{2}}\right) ,
	\label{tdkapp}
\end{displaymath}
which grows rapidly with increasing top mass.

If there are only three generations of quarks, so that $V_{tb}$ has a magnitude close 
to unity, then for a top-quark mass of 175\gevcc the partial width is
\begin{displaymath}
	\Gamma(t \rightarrow bW^+) \approx 1.55\gev,
	\label{eqn:twidth}
\end{displaymath} which corresponds to a top lifetime $\tau_t \approx 0.4 
\times 10^{-24}\s$, or 0.4 yoctosecond.  

The confining effects of the strong interaction act on a time scale 
of a few yoctoseconds set by $1/\hbox{the scale energy of quantum 
chromodynamics}$, $\Lambda_{\mathrm{QCD}}$.
This means that a top quark decays long before it can be hadronized.
There will be no discrete lines in toponium ($t\bar{t}$) spectroscopy, and indeed 
no dressed hadronic states containing top.  Accordingly, the characteristics of top 
production and of the hadrons accompanying top in phase space should be 
calculable in perturbative QCD \cite{pqcd}.  In top decay, we  
see the decay of an isolated quark, rather than the decay of a quark bound 
in a hadron. 
\vspace{12pt}
\noindent\HRule
\newpage
	\begin{table}[tbp]
		\caption{Channels studied in the search for the reaction $\bar{p}p 
		\rightarrow t\bar{t}+\hbox{ anything}$.  Those in parentheses have 
		not been exploited in experiments.  All but the 4 jets 
		$b\bar{b}$ mode must have significant ``missing'' transverse energy, carried 
		away by the neutrino(s) in the leptonic decay of the $W$ boson(s).}\vspace{12pt}
		\onepointfive
\begin{center}
		\begin{tabular}{cc}
			\hline
			Channel & Branching Fraction  \\
			\hline
			$e^{+}e^{-}b\bar{b}\met$ & 1/81  \\
			$\mu^{+}\mu^{-}b\bar{b}\met$ & 1/81  \\
			($\tau^{+}\tau^{-}b\bar{b}\met$ & 1/81)  \\[6pt]
			$e^{\pm}\mu^{\mp}b\bar{b}\met$ & 2/81  \\
			($e^{\pm}\tau^{\mp}b\bar{b}\met$ & 2/81)  \\
			($\mu^{\pm}\tau^{\mp}b\bar{b}\met$ & 2/81)  \\[6pt]
			$e^{\pm}\hbox{ jets }b\bar{b}\met$ & 12/81  \\
			$\mu^{\pm}\hbox{ jets }b\bar{b}\met$ & 12/81  \\
			($\tau^{\pm}\hbox{ jets }b\bar{b}\met$ & 12/81)  \\[6pt]
			$4\hbox{ jets }b\bar{b}$ & 36/81 \\
			\hline
		\end{tabular}
\end{center}
		\protect\label{canaux}
	\end{table}
	\vspace{72pt}
	\newpage
\newpage
\newpage
\newpage
\newpage


\begin{thebibliography}{19}
\bibitem{tatsu} The standard-model prediction for $M_{W}$ for 
	specified values of $m_{t}$ and $M_{H}$ was calculated using \textsf{ZFITTER 
	4.8} (D. Bardin, \etal, CERN--TH.6443/92).  The other input 
	parameters are $\alpha(M_{Z})=1/128.89$ and $\alpha_{s}(M_{Z}) = 
	0.123$.  I thank Tatsu Takeuchi for providing this technology.
\bibitem{su5} All the important features emerge in an $SU(5)$ unified 
theory that contains the standard-model gauge group $SU(3)_{c}\otimes 
SU(2)_{L} \otimes U(1)_{Y}$.  The final result is more general.
\end{thebibliography}
\end{document}